\documentclass[aps,showpacs,twocolumn]{revtex4}
\usepackage{amsfonts}
\usepackage{amsmath}
\usepackage{amssymb}
\usepackage{graphicx}
\usepackage{bm}

\input{tcilatex}

\begin{document}

\title{Thermocharge of a hot spot in an electrolyte solution}
\author{Arghya Majee and Alois W\"{u}rger}
\affiliation{Laboratoire Ondes et Mati\`{e}re d'Aquitaine, Universit\'{e} Bordeaux 1 \&
CNRS, 351 cours de la Lib\'{e}ration, 33405 Talence, France}
\pacs{66.10.C, 82.70.-y,47.57.J-}

\begin{abstract}
We discuss the thermoelectric properties of a locally heated micron-size
volume in an electrolyte solution. We find that such a hot spot carries a
net charge $Q$ which, for an excess temperature of 10 K, may attain hundreds
of elementary charges. The corresponding Seebeck electric field $E$
increases linearly with the radius $r$ inside the heated area, then goes
through a maximum, and decays as $1/r^{2}$ at larger distances. Our results
could be relevant for optothermal actuation of electrolytes and colloidal
suspensions.
\end{abstract}

\maketitle

\section{Introduction}

Heating a micron-size domain by a focussed laser beam has become a standard
technique of optothermal actuation in microfluidics. Thus a DNA trap was
realized through a\ thermal barrier in a microchannel \cite{Duh06a} and the
droplet size of a thermocapillary valve has been shown to be sensitive to
the Marangoni effect \cite{Bar07}. In a thin aqueous film, Soret-driven
depletion of polymers from a hot spot can be used for confining 100 nm
colloidal beads to a micro-domain \cite{Jia09}, or for separating solutes
like RNA by size \cite{Mae11}. As a recent biotechnological application,
protein interactions in biological liquids were studied by microscale
thermophoresis \cite{Wie10}.

The underlying thermal forces depend on the applied temperature gradient
through various mechanisms that are not always easily separated \cite%
{Wie04,Pia08,Wue10}. The common physical picture relies on local effects
such as thermoosmosis around solute particles \cite{Ruc81}, or Marangoni
forces along fluid interfaces \cite{Mur05}; molecular-dynamics simulations
suggest that even the molecular orientation, in both polar and non-polar
liquids, is sensitive to a temperature gradient \cite{Bre08,Roe12}. In
recent years it has become clear that, for charged systems in an electrolyte
solution, the thermoelectric or Seebeck effect provides a non-local driving
force that presents surprising properties.

In a non-uniform temperature, positive and negative salt ions have a
tendency to migrate in opposite directions, thus giving rise to a
thermopotential between the hot and cold boundaries of the sample \cite%
{Eas28,Gut49,Aga89}, and to a thermoelectric field $E=S\nabla T$ that is
proportional to the temperature gradient.\ In a colloidal suspension, this
field drives the solute particles at a velocity $u=\mu E$, where $\mu $ is
the Helmholtz-Smoluchowski electrophoretic mobility. The Seebeck coefficient 
$S$ depends on the electrolyte and may take either sign; the resulting
reversal of the drift velocity has been observed experimentally for
colloidal particles \cite{Put05} and SDS micelles \cite{Vig10}, upon
replacing NaCl ($S>0$) by sodium hydroxide NaOH ($S<0$). A detailed
comparison showed that the Seebeck effect dominates the thermoosmotic
pressure of mobile ions in the electric double layer of the colloid \cite%
{Wue08}. Moreover, it was found \cite{Maj11} that the thermoelectric effect
is at the origin of the observed dependence of the mobility on the colloidal
volume fraction \cite{Gho09}, and on the molecular weight of
polyelectrolytes \cite{Iac06} and DNA \cite{Duh06}. The recently reported
non-uniform variation of the Soret coefficient with the ionic strength \cite%
{Esl12} is characteristic for the electrophoretic mobility \cite{OBr78,Ant97}
and thus confirm the relevance of the Seebeck effect for the colloidal
motion. The examples cited so far concern a one-dimensional geometry, where
a constant temperature gradient arises from heating one side of the sample
and cooling the opposite side \cite{Lue12,Lue12a}. Because of the
macroscopic volume, boundary effects and in particular the surface charges
at the cold and hot walls are negligible.

\begin{figure}[b]
\includegraphics[width=\columnwidth]{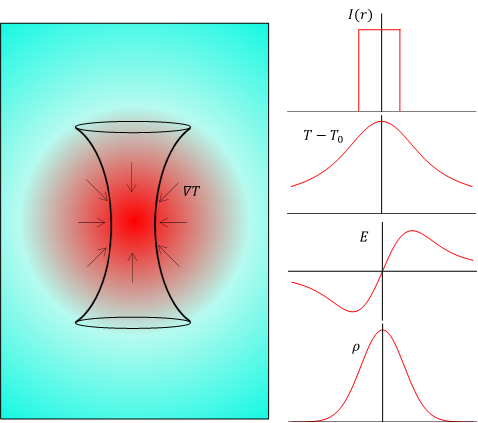}
\caption{Schematic view of a micron-size volume heated by absorption of a
focussed laser beam in an electrolyte solution. The right panel compares the
heating power $I(r)$ with the excess temperature $T-T_{0}$, the
thermoelectric field $E$, and the net charge density $\protect\rho $. }
\end{figure}

The present paper studies the thermoelectric properties of a locally heated
micron-sized volume in a bulk liquid, and discusses the ensuing colloidal
transport. The three-dimensional geometry results in a non-uniform
thermoelectric field and a net charge that is concentrated in the hot spot.
The present work is complementary to our recent study of the thermoelectric
properties of a heated colloidal particle \cite{Maj12}, where the spatial
separation of the heated volume (inside the bead) and the area where mobile
ions are present (outside) significantly simplified the problem.

Here we consider the thermoelectric properties of a hot spot in a bulk
liquid. The simple example of a square absorption profile is shown in Fig.
1. The excess temperature is a smooth function within the heated area and
decays with the inverse distance outside. Thermodiffusion of the mobile ions
induces a thermoelectric field that follows roughly the temperature
gradient, and is related to a net thermocharge confined within the heated
area. Since the salt ions diffuse within the waist of the focussed laser
beam, the electrostatic and thermal equations have to be solved
simultaneously. In Section 2 we present the general solution of the
electrolyte Seebeck effect in a non-uniform temperature. The case of a
Gaussian profile of the laser intensity is discussed in Section 3, and
simplified for the most relevant situation where the Debye length is small
as compared to the size of the heated area. Section 4 is devoted to
colloidal transport. In Sects. 5 and 6 we discuss the validity of the
approximations made and summarize our main results.

\section{Electrolyte Seebeck effect}

Suppose we have an electrolyte solution in a container and we are heating a
small region of this solution by a focused laser beam.\ The absorbed power
density $\beta I$ is determined by the laser intensity $I(\mathbf{r})$ and
the optical absorption coefficient $\beta $. Then the temperature profile is
solution of the stationary limit of Fourier's equation $\kappa \mathbf{%
\nabla }^{2}T+\beta I=0$, where $\kappa $\ is the heat conductivity. Its
formal solution reads 
\begin{equation}
T(\mathbf{r})-T_{0}=\frac{\beta }{\kappa }\int dV^{\prime }\frac{I(\mathbf{r}%
^{\prime })}{|\mathbf{r-r}^{\prime }|}.  \label{10}
\end{equation}%
Throughout this paper we consider an isotropic intensity, though in
experiments the transverse and longitudinal dimensions of the focus volume
in general differ from each other.

The bell-shaped temperature profile results in a thermal gradient that is
zero at the center of the heated spot, takes a maximum at its border, and
then decays with distance as $1/r^{2}$. As a consequence of this temperature
gradient, the mobile ions of the electrolyte solution will start moving. We
briefly present the underlying physical mechanisms and then discuss the
steady-state thermoelectric field.

\subsection{Ion thermodiffusion}

We consider a monovalent electrolyte solution of ionic strength $n_{0}$ and
non-uniform temperature $T$. Then the ion currents consist of three terms 
\cite{Gut49,Aga89,Wue10},%
\begin{equation}
\mathbf{J}_{\pm }=-D_{\pm }\left( \mathbf{\nabla }n_{\pm }+2\alpha _{\pm
}n_{\pm }\frac{\mathbf{\nabla }T}{T}\right) \pm \mu _{\pm }n_{\pm }e\mathbf{E%
},  \label{45}
\end{equation}%
where the first one corresponds to gradient diffusion with Einstein
coefficients $D_{\pm }$,\ the second accounts for thermal diffusion with the
dimensionless Soret parameters $\alpha _{\pm }$, and the last term describes
electrophoresis in the electric field $\mathbf{E}$ with the H\"{u}ckel
mobility $\mu _{\pm }$ of monovalent ions. For ions of radius much smaller
than the Debye length, these coefficients are related through \cite{Hie97} 
\begin{equation}
D_{\pm }=\mu _{\pm }k_{B}T.  \label{43}
\end{equation}

Eq. (\ref{45}), which arises in Onsager's theory for irreversible processes,
describes various non-equilibrium situations, depending whether the
concentration gradient, the temperature gradient, or the electric field are
taken as the external driving force \cite{Gro62}.\ As a well-known example,
we mention the case of an external salinity gradient $\mathbf{\nabla }n_{\pm
}\neq 0$ with uniform temperature $\mathbf{\nabla }T=0$: Unlike Einstein
coefficients of positive and negative ions lead to a diffusion electric
field $E\propto D_{+}-D_{-}$ and a \textquotedblleft chemiphoretic
effect\textquotedblright\ \cite{And89}.

Here we are interested in the Seebeck effect, which establishes the relation
between an applied temperature gradient and the resulting thermoelectric
field $E$. As source strength we have introduced the ionic Soret parameters $%
\alpha _{\pm }$ that are given by the ratio of thermodiffusion and diffusion
coefficients. The above definition implies that ions with positive $\alpha
_{\pm }$ migrate to the cold. In view of the early literature on non-uniform
salt solutions, these parameters may be considered as a dimensionless
measure of Eastman's \textquotedblleft entropy of
transfer\textquotedblright\ $S_{\pm }^{\ast }=2\alpha _{\pm }k_{B}$ \cite%
{Eas28}\ and Agar's \textquotedblleft heat of transport\textquotedblright\ $%
Q_{\pm }^{\ast }=2\alpha _{\pm }k_{B}T$ \cite{Aga89}. In physical terms, the 
$\alpha _{\pm }$ are determined by the ionic solvation free energy, which is
a complex function of electrostatic and hydration forces; at present there
is no satisfactory theoretical description for these parameters. From the
osmotic pressure of an ideal gas one obtains $\alpha _{\pm }=1$.\ For common
monovalent salt ions these numbers vary between 0 and 1 \cite{Wue08}; larger
values occur for molecular ions containing hydrogen.

\subsection{Steady state}

We discard transients due to an initial out-of equilibrium state. The steady
state is achieved if different contributions to the ion currents (\ref{45})
cancel each other, 
\begin{equation}
\mathbf{J}_{\pm }=0.  \label{46}
\end{equation}%
We briefly discuss the case where the thermodiffusion parameters are
identical $\alpha _{+}=\alpha _{-}$. Then the electric field vanishes, and
cations and anions show the same concentration modulation $n_{\pm }=n_{0}$.
The non-uniform salinity is described by $\mathbf{\nabla }n_{0}=-n_{0}S_{T}%
\mathbf{\nabla }T$ with the salt Soret coefficient $S_{T}=(\alpha
_{+}+\alpha _{-})/T$. Note that this \textquotedblleft Soret
equilibrium\textquotedblright\ involves diffusion and thermodiffusion
contributions to (\ref{45}) only, and thus implies $E=0$.

Now we turn to the electrolyte Seebeck effect that arises from the
difference $\alpha _{+}-\alpha _{-}$ of the single-ion Soret parameters.
Then the steady state condition (\ref{46}) cannot be achieved with the same
concentration profile for positive and negative ions. In other words, there
is a finite charge density 
\begin{equation*}
\rho =e(n_{+}-n_{-}),
\end{equation*}%
which in turn is related to an electric field. Mechanical equlibrium is
obtained where the thermodynamic and Coulomb forces cancel each other. The
steady-state electric field and charge distribution are related through
Gauss' law 
\begin{equation}
\mathbf{\nabla }\cdot \mathbf{E}=\rho /\varepsilon .  \label{47}
\end{equation}

\subsection{Linearization}

The conditions (\ref{46}) and (\ref{47}) determine the stationary
thermoelectric properties of the heated spot. Subtracting the equations $%
\mathbf{J}_{\pm }/D_{\pm }=0$ from each other, and introducing the gradient
of charge density $\mathbf{\nabla }\rho $ we get 
\begin{equation}
\mathbf{\nabla }\rho +2e(n_{+}\alpha _{+}-n_{-}\alpha _{-})\frac{\mathbf{%
\nabla }T}{T}-(n_{+}+n_{-})\frac{e^{2}\mathbf{E}}{k_{B}T}=0.  \label{44}
\end{equation}%
This intricate non-linear relation for the quantities $n_{\pm }$ simplifies
significantly when resorting to a linearization approximation in terms of
the out-of-equilibrium quantities, that is, the excess temperature, and the
resulting density changes and thermoelectric field. This amounts to
replacing, in the coefficients of $\mathbf{\nabla }T$ and $\mathbf{E}$, the
temperature $T$ and the ion densities $n_{\pm }$ with their bulk mean values 
$T_{0}$ and $n_{0}$. Note that we keep the density gradient in the first
term.

Thus the linearization approximation does not require small gradients $%
\mathbf{\nabla }n_{\pm }$ but relies on the weaker conditions 
\begin{equation}
\delta T\ll T_{0},\ \ \ \ \ \ \ |n_{\pm }-n_{0}|\ll n_{0}.  \label{20}
\end{equation}%
Retaining the gradient $\mathbf{\nabla }\rho $ but neglecting corrections in
the excess temperature $\delta T/T_{0}$ and the modulation of ion densities $%
n_{\pm }/n_{0}-1$, Eq. (\ref{44}) simplifies to 
\begin{equation}
\mathbf{\nabla }\rho +2en_{0}(\alpha _{+}-\alpha _{-})\frac{\mathbf{\nabla }T%
}{T_{0}}-2n_{0}\frac{e^{2}\mathbf{E}}{k_{B}T_{0}}=0.  \label{44a}
\end{equation}%
In the following we determine the Seebeck field $E$ from a set of two
coupled equations, which are Gauss' law (\ref{47}) and the linearized
thermoelectric relation (\ref{44a}).\ 

\subsection{Thermoelectric field}

Defining the Debye length $\lambda =(\varepsilon
k_{B}T_{0}/2e^{2}n_{0})^{1/2}$ and the Seebeck coefficient 
\begin{equation*}
S=\frac{k_{B}}{e}(\alpha _{+}-\alpha _{-}),
\end{equation*}%
we then obtain the equation for the electric field 
\begin{equation*}
\mathbf{E}=\lambda ^{2}\frac{\mathbf{\nabla }\rho }{\varepsilon }+S\mathbf{%
\nabla }T.
\end{equation*}%
This form shows that the electric field consists of two contributions: one
is related to the gradient of the charge density and the other one is
proportional to the temperature gradient. Inserting Gauss' law in the first
one, we get an inhomogeneous linear equation for the electric field:%
\begin{equation}
\lambda ^{2}\mathbf{\nabla (\nabla }\cdot \mathbf{E)}-\mathbf{E}+S\mathbf{%
\nabla }T=0.  \label{49}
\end{equation}

In order to avoid the vector derivatives, we insert $\mathbf{E}=-\mathbf{%
\nabla }\varphi $ and obtain the corresponding equation for the
electrostatic potential,\ 
\begin{equation}
\lambda ^{2}\mathbf{\nabla }^{2}\varphi -\varphi =S(T-T_{0}),  \label{11a}
\end{equation}%
where the constant $T_{0}$ gives the temperature far from the heated spot.
This is identical to the usual Debye-H\"{u}ckel equation, albeit with the
source field $S(T-T_{0})$. Since we consider an infinite bulk liquid without
internal boundaries, the formal solution of (\ref{11a}) is readily obtained
as the convolution of the screened propagator $g(\mathbf{r}^{\prime
})=e^{-r^{\prime }/\lambda }/4\pi r^{\prime }\lambda ^{2}$ with the
temperature profile, 
\begin{equation}
\varphi (\mathbf{r})=-S\int dV^{\prime }g(\mathbf{r}^{\prime })(T(\mathbf{r-r%
}^{\prime })-T_{0}).  \label{11}
\end{equation}%
The thermoelectric field is calculated by taking the gradient, 
\begin{equation}
\mathbf{E}(\mathbf{r})=S\int dV^{\prime }g(\mathbf{r}^{\prime })\mathbf{%
\nabla }T(\mathbf{r-r}^{\prime }).  \label{12}
\end{equation}%
\begin{table}[tbp]
\caption{Seebeck coefficient $S$ for NaCl, HCl, NaOH, and tetrabutylammonium
nitrate (TBAN) in aqueous solution \protect\cite{Aga89,Sok06,Bon11}. The
Seebeck coefficient is given by the difference of the \ reduced single-ion
Soret coefficients, $S=(k_{B}/e)(\protect\alpha _{+}-\protect\alpha _{-})$.
The dimensionless parameters $\protect\alpha _{\pm }$ are related to the
\textquotedblleft heat of transfer\textquotedblright\ $Q_{\pm }^{\ast
}=2k_{B}\protect\alpha _{\pm }$. For comparison, the Seebeck coefficient of
most simple metals is of the order of a few $\protect\mu $V/K.}
\ \ \ \ \ \ \ \ \ 
\begin{tabular}{|l|c|c|c|c|}
\hline
Salt & NaCl & NaOH & HCl & TBAN \\ \hline
$S$ (mV/K) & $0.05$ & $-0.22$ & $0.21$ & $1.0$ \\ \hline
$\alpha _{+}-\alpha _{-}$ & $0.6$ & $-2.7$ & $2.6$ & $12$ \\ \hline
\end{tabular}%
\end{table}

The thermoelectric field originates from the charge density $\rho
=\varepsilon \mathbf{\nabla }\cdot \mathbf{E}$ accumulated in the heated
spot \ by the thermodiffusion current. The total charge is obtained by
integrating over volume, 
\begin{equation*}
Q=\int dV\rho .
\end{equation*}%
Because of the overall neutrality, there is a countercharge $-Q$ at infinity
or, in the case of a finite system, at the sample container.

\subsection{Smooth temperature profile}

The size $a$ of the heated spot is at least of the order of microns, and
more often of tens of microns. Except for very weak electrolytes, this is
significantly larger than the Debye length $\lambda $; thus to leading order
in $\lambda /a$, the function $g(\mathbf{r}^{\prime })$ can be replaced with
Dirac's delta peak $\delta (\mathbf{r}^{\prime })$. More generally, this
approximation is valid as long as the temperature profile varies little over
one Debye length.\ 

Then the thermopotential is proportional to the excess temperature, $\varphi
_{0}(\mathbf{r})=-S(T(\mathbf{r})-T_{0})$, and the thermoelectric field is
given by the temperature gradient, 
\begin{equation}
\mathbf{E}_{0}=S\mathbf{\nabla }T.  \label{14}
\end{equation}%
Inserting this in Gauss' law (\ref{47}) and using Fourier's equation for
heat conduction, we find that the charge density is proportional to the
laser intensity,%
\begin{equation}
\rho _{0}(\mathbf{r})=\frac{\beta \varepsilon S}{\kappa }I(\mathbf{r}).
\label{15}
\end{equation}%
This relation is not surprising but follows directly from the above
expression for the thermopotential $\varphi _{0}(\mathbf{r})$. As a
consequence, the charge $Q$ accumulated in the heated spot is proportional
to the total absorbed power, 
\begin{equation}
Q=\frac{\beta \varepsilon S}{\kappa }\int dVI(\mathbf{r}).  \label{16}
\end{equation}

\section{Gaussian heating profile}

In many cases the intensity profile of the absorbed laser light is well
approximated by a Gaussian of maximum $I_{0}$ and width $a$,%
\begin{equation*}
I(\mathbf{r})=I_{0}e^{-r^{2}/a^{2}}.
\end{equation*}%
Then the formal expression (\ref{10}) for the temperature field is readily
integrated, 
\begin{equation}
T(\mathbf{r})-T_{0}=\frac{\sqrt{\pi }\beta I_{0}}{4\kappa r}\func{erf}\left( 
\frac{r}{a}\right) ,  \label{57}
\end{equation}%
with Gauss' error function%
\begin{equation*}
\func{erf}\left( x\right) =\frac{2}{\sqrt{\pi }}\int_{0}^{x}dte^{-t^{2}}.
\end{equation*}%
For small $x$ this function increases linearly and rapidly tends toward the
limiting value $\func{erf}\left( \infty \right) =1$ for $x>1$. $T-T_{0}$
becomes constant at the center of the heated spot, whereas at distances well
beyond the beam waist $a$, the excess temperature decreases as $1/r$.

\subsection{Electric field and thermocharge}

The complete expression for the thermoelectric field is obtained by solving (%
\ref{11a}) and then taking the gradient $E=-\nabla \phi $. We thus obtain

\begin{eqnarray}
E &=&-S\delta T\frac{a}{r^{2}}\left[ \func{erf}\left( \frac{r}{a}\right) -%
\frac{1}{2}e^{a^{2}/4\lambda ^{2}}\right. \times  \notag \\
&&\ \ \times \sum_{\pm }\left( \frac{r}{\lambda }\mp 1\right) e^{\pm
r/\lambda }\func{erf}\mathrm{c}\left( \frac{a}{2\lambda }\pm \frac{r}{a}%
\right) ,  \label{61}
\end{eqnarray}%
with the complementary error function $\func{erf}\mathrm{c}(x)=1-\func{erf}%
(x)$. In Fig. 2 we plot the thermoelectric field $E$ in units of $S\delta
T/a $ as a function of the reduced distance $r/a$, for different values of
the Debye screening length. The five curves are obtained for $\lambda
/a=3;1;0.3;0.1;0$. The upper one, ($\lambda /a=0$) corresponds to $S\nabla T$%
, that is, the corrections in (\ref{61}) vanish. At large distances $r\gg
a+\lambda $, the electric field is independent of $\lambda $, and all curves
converge toward $S\nabla T$. Within a volume of radius $a+\lambda $, a
significant reduction occurs; as the ratio $\lambda /a$ increases, the field 
$E$ becomes smaller than $S\nabla T$.

The charge density is calculated from Gauss' law. From (\ref{61}) we obtain%
\begin{equation}
\rho =\frac{\varepsilon S\delta Ta}{2\lambda ^{2}r}e^{a^{2}/4\lambda
^{2}}\sum_{\pm }\pm e^{\pm r/\lambda }\func{erf}\mathrm{c}\left( \frac{a}{%
2\lambda }\pm \frac{r}{a}\right) .  \label{70}
\end{equation}%
This expression decays exponentially. Total charge ($Q$) accumulated within
and very close to the heated region can be calculate by integrating the
charge density and this gives%
\begin{equation}
Q=-4\pi a\varepsilon S\delta T_{0}=-e(\alpha _{+}-\alpha _{-})\frac{a}{\ell
_{B}}\frac{\delta T}{T_{0}}.  \label{66}
\end{equation}%
In the second equality we have used the definition of Bjerrum length $\ell
_{B}=e^{2}/4\pi \varepsilon k_{B}T_{0}$ and expressed the Seebeck
coefficient through the ion Soret parameters $\alpha _{\pm }$. The net
charge accumulated is proportional to the excess temperature and to the
radius of the heated spot.

\begin{figure}[b]
\includegraphics[width=\columnwidth]{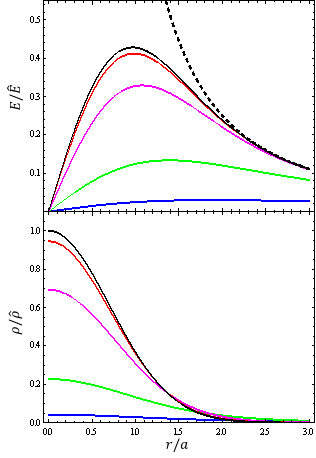}
\caption{Upper panel: Thermoelectric field $E$ in units of $\widehat{E}=-S%
\protect\delta T/a $ as a function of the reduced radial distance $r/a$. The
solid curves are calculated for a Gaussian heating profile; from above they
show Eq. (\protect\ref{61}) with $\protect\lambda /a=0;0.1;0.3;1;3$. Thus
the top most solid curve corresponds to Eq. (\protect\ref{62}). The dashed
line gives the behavior of the electric field of a point charge,
proportional to $1/r^{2}$. For a hot spot in a NaCl solution of radius $a=1%
\protect\mu $m and excess temperature $\protect\delta T=10$ K, the scale
factor takes the value $\widehat{E}=-0.5$ kV/m. Lower Panel: Charge density $%
\protect\rho $ for Gaussian laser intensity in units of $\widehat{\protect%
\rho }=-4\protect\varepsilon S\protect\delta T/\protect\sqrt{\protect\pi }%
a^{2}$. The curves are calculated from Eq. (\protect\ref{70}) with the above
parameters, resulting in $\widehat{\protect\rho }=-5$ e/$\protect\mu $m$^{3}$%
. Except for the upper one, which corresponds to Eq. (\protect\ref{65}),
these curves are not Gaussians; larger values of the ratio $\protect\lambda %
/a$ flatten the charge density which approximately covers the domain within $%
a+\protect\lambda $. }
\end{figure}

\subsection{Limiting case $\protect\lambda /a\rightarrow 0$}

Most real systems correspond to the limiting case $\lambda \ll a$. Indeed,
the size of the heated spot is at least several microns, whereas the Debye
length takes values between 1 and 100 nanometers. Then a power series for $%
\func{erf}(x)$ in terms of $1/x$ provides a useful approximation, 
\begin{equation*}
\func{erf}(x)=1-\frac{e^{-x^{2}}}{\sqrt{\pi }x}\left( 1-\frac{1}{2x^{2}}%
+...\right) .
\end{equation*}%
Inserting in the above form for $E$ and retaining the leading terms only, we
have 
\begin{equation}
E_{0}=-S\delta T\frac{a}{r^{2}}\left( \func{erf}\left( \frac{r}{a}\right) -%
\frac{2}{\sqrt{\pi }}\frac{r}{a}e^{-r^{2}/a^{2}}\right) =S\nabla T.
\label{62}
\end{equation}%
The second equality is readily obtained from the temperature field (\ref{57}%
),\ in accordance with the approximation in the general case defined in (\ref%
{14}). By the same token the charge density (\ref{70}) simplifies to 
\begin{equation}
\rho _{0}=-\frac{4\varepsilon S\delta T}{\sqrt{\pi }a^{2}}e^{-r^{2}/a^{2}}=%
\widehat{\rho }e^{-r^{2}/a^{2}}.  \label{65}
\end{equation}%
As expected from (\ref{15}), this expression is proportional to the laser
intensity. The second equality defines the scale factor for the plots in
Fig. 2.

Equations (\ref{62}) and (\ref{65}) are plotted in Fig. 2 with the label $%
\lambda /a=0$. In the case of finite but small Debye length, these relations
provide a very good approximation both inside and outside the heated spot;
for $\lambda /a<0.03$, the error is smaller than the line width in Fig.\ 2.\
For a micron-size particle this condition is met for $\lambda <30$ nm, that
is for an electrolyte strength of at least $10^{-4}$ M/l.

\section{Colloidal transport}

\subsection{Drift velocity}

In recent years local heating with an infrared laser has widely used for
confining or sieving macromolecules or colloidal particles in aqueous
solution \cite{Duh06a,Jia09,Mae11,Wie10}. These works rely on the drift
velocity $u$ imposed by the temperature gradient. Various physical
mechanisms have been discussed \cite{Wue10}; in the present context the most
relevant are 
\begin{equation}
u=-\frac{\varepsilon \zeta ^{2}}{3\eta }\frac{\nabla T}{T}+\frac{\varepsilon
\zeta }{\eta }E,  \label{74}
\end{equation}%
where $\zeta $ is the surface potential, $\varepsilon $ the solvent
permittivity, and $\eta $\ the viscosity.

In a macroscopic system, the thermoelectric field is strictly proportional
to the temperature gradient, $E=S\nabla T$, such that (\ref{74}) can be
rewritten as 
\begin{equation}
u_{0}=-D_{T}\nabla T,\ \ \ \ D_{T}=\frac{\varepsilon \zeta ^{2}}{3\eta T}-%
\frac{\varepsilon \zeta }{\eta }S,  \label{76}
\end{equation}%
where the thermophoretic mobility $D_{T}$ is taken as a constant \cite{Wue10}%
. The first term in $D_{T}$ arises from the thermoosmotic salt-ion flow
around a collodial bead; it was first derived by Ruckenstein \cite{Ruc81}
and drives the solute particle to the cold. The second term in (\ref{74})
accounts for electrophoresis in the Seebeck field $E$. In the last few years
it has become clear that this thermoelectric effect contributes
significantly to collodial thermophoresis and in many cases is even dominant 
\cite{Put05,Vig10,Wue08}. The velocity may even change its sign with the
Seebeck coefficient: This means that negatively charged colloidal particles
or SDS micelles move to the warm in NaOH ($S<0$) and to the cold in NaCl ($%
S>0$). As a consequence, the colloid accumulates or is depleted,
respectively.

\subsection{Colloidal accumulation at finite radius}

For a micron-size hot spot in a very weak electrolyte, the Debye length is
of the same order of magnitude of the spot size. Then the thermoelectric
properties do not reduce to the macroscopic Seebeck field $E_{0}=S\nabla T$
but result in a significantly more involved relation (\ref{61}) between $E$
and $\nabla T$. As the most striking feature of Fig.\ 2, moderate or large
values of $\lambda /a$ strongly reduce the thermoelectric field in the
heated area, yet are of no effect at larger distances.

\begin{figure}[b]
\includegraphics[width=\columnwidth]{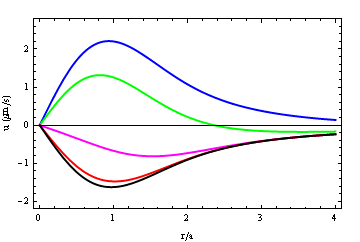}
\caption{Spatial variation of the drift velocity (\protect\ref%
{74}) for different values of the ratio $\protect\lambda /a$. From below the
curves are calculated with $\protect\lambda /a=0;0.1;0.3;1;3$. }
\end{figure}

In Fig. 3 we plot the spatial variation of the drift velocity $u$ of a
colloidal solute in NaOH solution, for different values of the ratio $%
\lambda /a$. Because of the negative Seebeck coefficient of NaOH, the
thermoelectric field is opposite to the temperature gradient. Thus for a
negatively charged solute, the two terms in (\ref{74}) carry opposite signs:
The Ruckenstein mechanism drives the solute toward the cold, whereas the
Seebeck contribution points to the origin of the heated spot. For small $%
\lambda /a$ and with the numbers of Table 1, the Seebeck term is dominant
everywhere, and the solute migrates to higher temperature $(u<0)$. As the
ratio $\lambda /a,$ increases, the field $E$ decreases below $E_{0}$ within
the heated area, yet remains unaffected at $r>a+\lambda $ where $E=E_{0}$.
As a consequence, the first term in (\ref{74}) dominates at at small
distances. Accordingly, the solute migrates outward at small $r$, and in the
opposite direction at larger distances; thus the solute accumulates where $u$
vanishes. A similar effect has been shown to result from the competition
between thermophoresis and depletion forces \cite{Mae11,Oda12}.

\begin{figure}[b]
\includegraphics[width=\columnwidth]{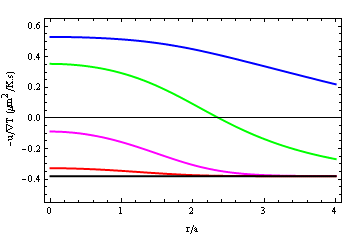}
\caption{ Spatial variation of the effective
thermophoretic mobiltiy $-u/\protect\nabla T$ for different values of the
ratio $\protect\lambda /a$. From below the curves are calculated with $%
\protect\lambda /a=0;0.1;0.3;1;3$. The lowest one corresponds to the
mobility $D_{T}$ defined in (\protect\ref{76})}
\end{figure}

In order to single out the relative variation of the thermoelectric field,
we plot in Fig. 4 the ratio of the velocity $u$ and the underlying
thermodynamic force $\nabla T$. The lowest curve is calculated with $\lambda
/a=0$, in other words with the Seebeck field $E_{0}=S\nabla T$; the constant
value $-u/\nabla T=D_{T}$ gives the usual thermophoretic mobility of the
colloid. As $\lambda /a$ increases, the ratio between $u$ and $\nabla T$ is
no longer constant, and the thermoelectric contribution is reduced within
the heated spot and in its immediate vicinity. For large $\lambda /a$ and at
short distances, the thermoelectric field is negligible and $-u/\nabla T$
tends toward the (positive) value of Ruckenstein contribution in (\ref{74}).
As the distance $r$ increases, the ratio $-u/\nabla T$ changes sign and
finally converges to the mobility $D_{T}$.

\subsection{Temperature dependence}

In the preceding discussion we have assumed that the coefficients of $\nabla
T$ in (\ref{74}) and (\ref{76}) are constant in the heated spot and beyond.
It is well known that the properties of both solute and solvent (viscosity,
permittivity, Seebeck parameter, $\zeta $-potential) vary with temperature,
in addition to the explicit appearance of $T$ in (\ref{74}). An estimate for
real systems shows the largest dependencies occur for the viscosity and
Seebeck coefficient; their logarithmic derivatives $d\ln \eta /dT$ and $d\ln
S/dT$ are of the order of 0.02 K$^{-1}$ \cite{Wue09}, which means that an
excess temperature of 10 K results in a change of about 20 percent. There
are however, several good reasons for discarding this variation when
calculating the drift velocity $u$.

(i) These corrections are formally similar to those neglected in the
linearization approximation of (\ref{44}). Retaining the former and
discarding the latter would be inconsistent. (ii) If the temperature
dependence of viscosity and permittivity are rather well known and thus
could be taken into account, this is not the case for the Seebeck
coefficient $S$ nor for the $\zeta $-potential: There is some evidence that
the ionic Soret parameters $\alpha _{\pm }$, and thus $S$, are correlated\
with the thermal expansivity of water \cite{Wue09}; unfortunately there are
little data, and a satisfactory theory for the Seebeck coefficient is
lacking so far. Regarding the surface potential $\zeta $, its complex
temperature dependence is due to the boundary-layer electrostatics and the
dissociation of the charged molecular units. In view of these poorly
understood dependencies, characterizing the variation of thermoelectric
properties with temperature would seem a rather difficult undertaking.

Though not insignificant, these variations are of little relevance for the
present discussion, which focusses on the order of magnitude of the
thermoelectric effect and on its sign. Thus the coefficients of $\nabla T$
in (\ref{74}) and (\ref{76}) are calculated with constant $\eta ,\varepsilon
,S,\zeta $, and $T$. This is why the mobility $D_{T}$, that is, the lowest
curve of Fig. 4, does not vary with the distance.

\subsection{Concentration dependence}

The drift velocity (\ref{74}) and (\ref{76}) has been calculated in the
dilute limit, where the Seebeck field $E$ is determined by the salt ions but
independent of the colloidal charges. Here a word is in order concerning the
back-reaction of the latter on the thermoelectric properties. In a recent
work \cite{Maj11} we have shown that for polyelectrolytes, collective
effects set at a volume fraction $\phi \sim \ell _{B}/R_{g}$, where $\ell
_{B}=7$ \AA\ is the Bjerrum length and $R_{g}$ the gyration radius. A
slightly more complex expression occurs for colloidal particles, because of
their cooperative diffusion.

Comparison with several experiments suggests that collective effects occur
at rather moderately dense suspensions, requiring to go beyond the
single-particle picture (\ref{74}). This is certainly the case in
accumulation experiments where thermophoresis contributes to colloidal
confinement \cite{Bra02}.

\section{Discussion}

\subsection{Linearization approximation}

The analytic results of this paper rely essentially on linearizing the
differential equation (\ref{44}) in terms of excess temperature and
modulation of ion densities. Though much higher values can be achieved
experimentally \cite{Rin10}, typical values for $\delta T$ are of the order
of tens of Kelvin, thus satisfying the first inequality in (\ref{20}).
Regarding the ion densities, a first estimate is obtained by noting that
their modulation is of the order $n_{\pm }-n_{0}\sim \alpha _{\pm
}n_{0}\delta T/T$; the parameters $\alpha _{\pm }$ taking values of the
order of unity \cite{Aga89}, a small excess temperature implies a small
change of the ion densities.

Since the thermally induced charge density $\rho $ is a central quantity, we
discuss in some detail the fraction $\rho /en_{0}$ of salt ions involved. We
start with the case of small Debye length $\lambda /a\ll 1$. Inserting the
definition $\lambda =(\varepsilon k_{B}T_{0}/2e^{2}n_{0})^{1/2}$ in (\ref{65}%
) we find up to a numerical constant 
\begin{equation}
\frac{\hat{\rho}}{en_{0}}=\frac{eS}{k_{B}}\frac{\lambda ^{2}}{a^{2}}\frac{%
\delta T}{T_{0}}.  \label{72}
\end{equation}%
According to the numbers of Table I, the first factor on the right-hand side
is of the order of unity, which just means $\alpha _{\pm }\sim 1$. Both of
the remaining factors is small however, resulting in $\hat{\rho}\ll en_{0}$.
As an example, we consider the parameters of Fig. 2 with $\hat{\rho}=-5e\
\mu $m$^{-3}$; for comparison, the ion density of a 1 mM/l electrolyte
solution, $n_{0}=6\times 10^{5}\ \mu $m$^{-3}$, is by five orders of
magnitude larger than the net charge density. For the case where the Debye
length exceeds the size of the hot spot, $\lambda >a$, one finds a relation
similar to (\ref{72}), albeit without the factor $\lambda ^{2}/a^{2}$.
Again, for typical excess temperatures $\delta T\sim 10$K, the net charge
density turns out to be by almost two orders of magnitude smaller than
salinity. Thus the inequality 
\begin{equation*}
\rho \ll en_{0}
\end{equation*}%
holds independently of the ratio $\lambda /a$. This means that thermal
charge separation involves only a small fraction of the ion density.

On the other hand, this inequaltiy implies the second relation of (\ref{20}%
). Writing the charge density as $\rho =e(n_{+}-n_{0})-e(n_{-}-n_{0})$ and
noting that it is dominated by the ion species with the larger heat of
transfer parameter $\alpha _{\pm }$, for example $\rho \approx
-e(n_{-}-n_{0})$ in the case of NaOH ($\alpha _{+}=0.7$ and $\alpha _{-}=3.4$%
), we find that $|\rho |\ll en_{0}$ implies the concentration change of both
cations and anions to be small. These estimates confirm the validity of the
linearization approximation for the fundamental equation (\ref{44}). Since
experimental studies of confinement, separation, and transport, hardly
exceed an excess temperature $\delta T$  of 15 K, the thermoelectric
properties should be well described by the present work. The smooth
functional behavior of Eq. (\ref{44}) suggests that our qualitative results
hold true even at much larger excess temperature $\delta T\sim T$, where $%
\hat{\rho}/en_{0}\sim \lambda ^{2}/a^{2}\ll 1$. 

\subsection{Thermocharge and Seebeck field}

The main results of the present work are the thermocharge $Q$ accumulated in
the heated spot, and the corresponding electric field $E$. According to Eq. (%
\ref{66}) the charge is proportional to the reduced Seebeck parameter $%
\alpha _{+}-\alpha _{-}$, to the size of the spot in units of the Bjerrum
length $\ell _{B}=0.7$ nm, and to the reduced excess temperature. For a spot
of $a=5\mu $m radius and excess temperature $\delta T=30$ K, the induced $Q$
takes a value of about thousand elementary charges.\ 

At distances well beyond the heated spot, $r\gg a$, the thermoelectric field
takes the form 
\begin{equation}
E=\frac{Q}{4\pi \varepsilon r^{2}}.  \label{73}
\end{equation}%
The variation with the inverse square of the distance, characteristic of an
unscreened field, highlights the fact that the heated spot carries a net
charge $Q$. The field takes its maximum value at $r\sim a$; with the above
parameters, it is of the order of kV/m.

The ratio of the Debye length $\lambda $ and the size of the hot spot $a$
turns to be an important parameter. For $\lambda \ll a$ we find that the
thermo-charge density $\rho $ is proportional to the heating power, as shown
in (\ref{15}) for the general case and in (\ref{65}) for a Gaussian heating
profile. A more complex situation occurs if $\lambda $ is not small as
compared to the size of the heated domain. The lower panel of Fig.\ 2 shows
that already at $\lambda /a=\frac{1}{3}$ the charge density in the center is
reduced by one third, and augmented at large distances accordingly. As a
consequence of this smearing out of the thermocharge, the electric field is
reduced within and in the vicinity of the heated area; at larger distances $%
E $ tends towards (\ref{73}).

\subsection{Thermally driven convection}

So far we have neglected convection induced in the liquid by the temperature
gradient. Because of the thermal expansion, the heated spot has a slightly
lower density.\ Thus gravity drives an upward convective flow, similar to
the Rayleigh-B\'{e}nard cells that develop in a horizontal layer of fluid
heated from below. An estimate of velocity is obtained by equilibrating the
buoyancy and Stokes drag forces \cite{Rus04}, 
\begin{equation*}
u_{c}\sim \beta \delta Tga^{2}/\nu ,
\end{equation*}%
with the thermal expansivity $\beta =2\times 10^{-4}$ K$^{-1}$ and kinematic
viscosity $\nu =10^{-6}$ m$^{2}$/s of water at room temperature. For a
heated spot of size $a=10\mu $m and excess temperature $\delta T=10$ K one
finds a convective velocity $u_{c}$ of about 10 $\mu $m/s.

An important question is whether or not the mobile charges are advected by
this flow. This is best answered in term of the P\'{e}clet number, 
\begin{equation*}
\mathrm{Pe}=\frac{u_{c}}{D/a},
\end{equation*}
that compares the convection velocity with diffusion over the characteristic
length $a$. With the Einstein coefficient of common ions $D=2\times 10^{-9}$
m$^{2}$/s and $a=10\mu $m, one finds $D/a=200$ $\mu $m/s and a P\'{e}clet
number of about 5\%. In physical terms this means that the salt ions diffuse
sufficiently rapidly, such that the charge profile $\rho $ and the
thermoelectric field $E$ are hardly affected by convection.

A different picture arises for\ a colloidal suspension. Because of the much
smaller Einstein coefficient, $D=2\times 10^{-12}$ m$^{2}$/s for 100 nm
beads, the P\'{e}clet number is larger than unity and diffusion is much
slower than advection \cite{Rus04}. The interplay between diffusion,
convection, and thermophoresis has been used in thermal traps and separation
devices \cite{Duh06a,Mae11}.

Finally we note that the present work deals with a stationary temperature
profile. Additional flows arise from time-dependent heating, as illustrated
by experiments using thermoviscous expansion \cite{Wei08}.

\subsection{Non-spherical geometries}

We conclude with a brief discussion of non-spherical geometries. In most
cases the laser intensity $I(\mathbf{r})$ has no spherical symmetry, that
is, it shows different profiles parallel and perpendicular to the beam axis,
and so does the excess temperature. Then (\ref{49}) and (\ref{11a}) do not
reduce to a single radial equation. Still, if the spatial variation of $T$
is slow on the scale of the Debye length, $\lambda |\nabla T|\ll \delta T$,
the thermoelectric field follows the temperature gradient according to the
macroscopic law $E_{0}=S\nabla T$. As an important result, Eqs. (\ref{14})
and (\ref{15}) are valid for any sufficienlty smooth heating profile; this
implies in particular the existence of the thermocharge (\ref{66}), with
some characteristic length $a$.

Thus the analysis of Sect. 2 is valid for any smooth temperature profile.
The results obtained in Sects 3 and 4, however, and in particular the
electric field (\ref{61}) and the thermophoretic velocity (\ref{76}), rely
on the spherical symmetry of both thermal and electric properties.
Significant deviations occur for a non-spherical heated spot that is not
much larger than the Debye length, or for an electrolyte that is confined by
a solid boundary. We briefly discuss the latter situation.

\begin{figure}[b]
\includegraphics[width=\columnwidth]{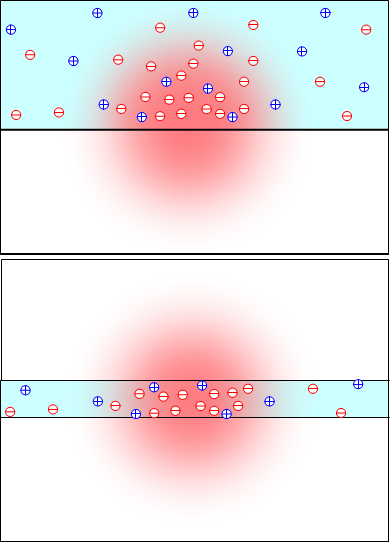}
\caption{Heated spot at a liquid-solid
boundary (upper panel) or in thin aqueous film (lower panel).}
\end{figure}

We consider a confined electrolyte solution, where the ratio of the
permittivities of water $\varepsilon $ and the insulating boundary $%
\varepsilon _{\text{in}}$ constitutes a most important parameter; for
typical materials one has $\varepsilon /\varepsilon _{\text{in}}\sim 30$.
The upper panel of Fig. 5 shows a heated spot at a liquid-solid or
liquid-liquid interface. Though this geometry does not allow for a general
solution of the thermoelectric equations, qualitative features are readily
obtained by comparing with results for a discrete surface charge $Q$ \cite%
{For02}. Estimating the ion pressure and the Maxwell stress tensor $\mathcal{%
T}$ in the framework of Debye-H\"{u}ckel theory, one finds a normal stress
component $\mathcal{T}\sim Q^{2}/\varepsilon _{\text{in}}a^{4}$ that pushes
the interface toward the electrolyte; the resulting force\ on the heated
spot, $F\sim Q^{2}/\varepsilon _{\text{in}}a^{2}$, is counterbalanced by the
opposite stress on the outside area \cite{For04}. This thermoelectric stress
could contribute to the temperature-driven forces on droplets that are used
for optothermal actuation in microfluidic devices \cite{Bar07}.

Even more striking effects are expected to occur in a locally heated thin
aqueous film, as illustrated in the lower panel of Fig. 5. The ratio of the
film thickness $h$ and the size $a$ of the hot spot provide an important
parameter for the thermoelectric equations. For $h\gg a$ one recovers the
bulk behavior studied in this paper, whereas a rather different picture
arises in the opposite case $h<a$. The permittivity jump at the interfaces
confines the electric field lines and results in a strongly non-uniform
relation between $E$ and $\nabla T$, similar to that discussed above for the
ratio $\lambda /a$. According to our preliminary results, this confinement
could be relevant for the thermophoretic mobility observed for thin-film
geometries \cite{Duh06,Mae11,Jia10}.

\section{Summary}

We briefly summarize our main results on the thermoelectric properties of a
locally heated spot in an electrolyte solution.

(i) For a smooth temperature profile, that varies little within one Debye
length, the Seebeck field (\ref{14}) and the thermocharge density (\ref{15})
are proportional to the temperature gradient and laser intensity,
respectively. These results does not rely on a particular geometry but hold
true in general.

(ii) The total charge (\ref{16}) is proportional to the integrated power.
With typical experimental parameters, $Q$ may attain thousand elementary
charges. 

(iii) Laser heating can be used for designing local electric fields of the
order of kV/m, without electrodes or other material in the electrolyte
solution. This thermoelectric field is not screened and, beyond the heated
spot, decays with the square of the inverse distanc\.{e}. 

(iv) In recent years, thermal gradients have been used for accumulating or
depleting colloidal solutes in a hot spot. The Seebeck field derived here
could contribute significantly to the thermodynamic forces observed
experimentally, and provide an qualitative picture for situations where
convection is of little relevance, for example in confined micron-size
films. Regarding thermoconvection, which is present in a bulk system, we
find that salt ions are hardly affected (small P\'{e}clet number), whereas\
the stationary state of colloidal solutes is to a large extent determined by
advection.\


\begin{thebibliography}{99}
\bibitem{Duh06a} S.\ Duhr and D.\ Braun, \textit{Phys. Rev. Lett.}, 2006, 
\textbf{97}, 038103.

\bibitem{Bar07} C.\ Baroud, J.-P. Delville, F. Gallaire and R. Wunenburger, 
\textit{Phys. Rev. E: Stat., Nonlinear, Soft Matter Phys.}, 2007,\ \textbf{%
75,} 046302.

\bibitem{Jia09} H.-R. Jiang, H.\ Wada, N. Yoshinaga and M. Sano, \textit{%
Phys. Rev. Lett.}, 2009, \textbf{102}, 208301.

\bibitem{Mae11} Y.T. Maeda, A. Buguin and A. Libchaber, \textit{Phys. Rev.
Lett.}, 2011, \textbf{107}, 038301.

\bibitem{Wie10} C.J.\ Wienken, Ph.\ Baaske, U.\ Rothbauer, D.\ Braun and S.\
Duhr, \textit{Nature Communications}, 2010, \textbf{1}, 100.

\bibitem{Wie04} S.\ Wiegand, J. Phys. Cond. Matt. \textbf{16}, 357 (2004)

\bibitem{Pia08} R. Piazza, \textit{Soft Matter}, 2008, \textbf{4}, 1740.

\bibitem{Wue10} A. W\"{u}rger, \textit{Rep. Prog. Phys.}, 2010, \textbf{73},
126601.

\bibitem{Ruc81} E. Ruckenstein, \textit{J. Coll. Interf. Sci.}, 1981, 
\textbf{83}, 77.

\bibitem{Mur05} R.M. Muruganathan, Z. Khattari and Th. M. Fischer, \textit{%
J.\ Phys. Chem. B}, 2005, \textbf{109}, 21772.

\bibitem{Bre08} F. Bresme, A. Lervik, D. Bedeaux, and S. Kjelstrup, \textit{%
Phys. Rev. Lett.}, 2008, \textbf{101}, 020602.

\bibitem{Roe12} F. R\"{o}mer, F. Bresme, J. Muscatello, D. Bedeaux and J.
Miguel Rub\i , \textit{Phys. Rev. Lett.}, 2012, \textbf{108}, 105901.

\bibitem{Eas28} E. D. Eastman, \textit{J. Am. Chem. Soc.}, 1928, \textbf{50}%
, 292-297.

\bibitem{Gut49} G. Guthrie, J. N. Wilson and V. Schomaker, \textit{J.\ Chem.
Phys.}, 1949,\ \textbf{17}, 310-313.

\bibitem{Aga89} J. N. Agar, C. Y. Mou and J.-L. Lin, \textit{J.\ Phys.\ Chem.%
}, 1989, \textbf{93}, 2079-2082.

\bibitem{Put05} S.A. Putnam and D.G.\ Cahill., \textit{Langmuir}, 2005, 
\textbf{21}, 5317.

\bibitem{Vig10} D. Vigolo, S.\ Buzzaccaro and R.\ Piazza, \textit{Langmuir},
2010, \textbf{26}, 7792.

\bibitem{Wue08} A. W\"{u}rger, \textit{Phys. Rev. Lett.}, 2008, \textbf{101}%
, 108302.

\bibitem{Maj11} A.\ Majee and A. W\"{u}rger, \textit{Phys. Rev. E: Stat.,
Nonlinear, Soft Matter Phys.}, 2011, \textbf{83}, 061403.

\bibitem{Gho09} N. Ghofraniha, G. Ruocco and C. Conti, \textit{Langmuir},
2009, \textbf{25}, 12495.

\bibitem{Iac06} S.\ Iacopini, R. Rusconi, and R. Piazza, \textit{Eur. Phys.
J. E}, 2006, \textbf{19}, 59-67.

\bibitem{Duh06} S.\ Duhr and D.\ Braun, \textit{Proc. Natl. Acad. Sci. USA},
2006, \textbf{103}, 19678-19682.

\bibitem{Esl12} K..\ Eslahian and M.\ Maskos, \textit{Colloids and Surfaces A%
}, 2012

\bibitem{OBr78} R. W. O'Brian and L. R. J. White, \textit{J. Chem.\ Soc.
Faraday Trans. II.}, 1978, \textbf{74}, 1607.

\bibitem{Ant97} M. Antonietti and L. Vorwerg, \textit{Colloid Polym Sci.}%
,1997,\textbf{\ 275}, 883.

\bibitem{Lue12} D.\ L\"{u}sebrink, M.\ Ripoll, J. Chem. Phys. 136, 084106
(2012)

\bibitem{Lue12a} D.\ L\"{u}sebrink, M.\ Yang, M.\ Ripoll, J. Phys.: Condens.
Matter \textbf{24}, 284132 (2012)

\bibitem{Maj12} A.\ Majee and A. W\"{u}rger, \textit{Phys. Rev. Lett.},
2012, \textbf{108}, 118301.

\bibitem{Hie97} P.C. Hiemenz, R.\ Rajagopalan, \textit{Principles of Colloid
and Surface Chemistry}, Dekker (1997)

\bibitem{And89} J.O.\ Anderson, \textit{Ann. Rev. Fluid Mech.}\ 1989, 
\textbf{21}, 61.\ \ 

\bibitem{Gro62} S.R. de Groot and P.\ Mazur, \textit{Non-equlibrium
thermodynamics}, North Holland Publishing, Amsterdam, 1962.

\bibitem{Sok06} V. N. Sokolov, L. P. Safonova and A. A. Pribochenko, \textit{%
J. Solution Chem.}, 2006, \textbf{35}, 1621.

\bibitem{Bon11} M.\ Bonetti, S.\ \ Nakamae, M.\ Roger and P.\ Guenoun, 
\textit{J.\ Chem.\ Phys.}, 2011, \textbf{134}, 114513.

\bibitem{Oda12} K. Odagiri, K. Seki and K. Kudo, \textit{Soft Matter} 2012, 
\textbf{8}, 6775-6781.

\bibitem{Wue09} A.\ W\"{u}rger, \textit{Langmuir }2009, \textbf{25},
6696-6701.

\bibitem{Bra02} D. Braun and A. Libchaber, \textit{Phys. Rev. Lett.}, 2002, 
\textbf{89}, 188103.

\bibitem{Rin10} D. Rings, R. Schachoff, M. Selmke, F. Cichos, K. Kroy, 
\textit{Phys. Rev. Lett. }2010, \textbf{105}, 090604.

\bibitem{Rus04} R. Rusconi, L. Isa, R. Piazza, \textit{J. Opt. Soc. Am.}
2004, \textbf{21}, 605

\bibitem{Wei08} F.M. Weinert, D.\ Braun, \textit{J. Appl. Phys.} 2008, 
\textbf{104}, 104701

\bibitem{For02} L. Foret, R. K\"{u}hn, A. W\"{u}rger, \textit{Phys. Rev.
Lett.} 2002, \textbf{89}, 156102.

\bibitem{For04} L. Foret, A. W\"{u}rger, \textit{Phys. Rev. Lett.} 2004, 
\textbf{92}, 058302.

\bibitem{Jia10} H.-R. Jiang, N. Yoshinaga, M. Sano, \textit{Phys. Rev. Lett.}%
, 2010, \textbf{105}, 268302.
\end{thebibliography}
\end{document}